\documentclass[10pt,journal,compsoc]{IEEEtran}

\ifCLASSOPTIONcompsoc
  \usepackage[nocompress]{cite}
\else
  \usepackage{cite}
\fi

\usepackage{amsmath}
\usepackage{amssymb}
\usepackage{xcolor}
\usepackage{colortbl}
\usepackage{float}
\usepackage{algorithm}
\usepackage{algorithmic}
\usepackage{graphicx}
\usepackage{subfig}

\newcommand{\R}{\mathbb{R}}

\begin{document}

\title{Utilising Deep Learning and Genome Wide Association Studies for Epistatic-Driven Preterm Birth Classification in African-American Women}

\author{\IEEEauthorblockN{Paul Fergus,
  Casimiro Curbelo Monta\~{n}ez,
  Basma Abdulaimma, Paulo Lisboa, and Carl Chalmers
}
\IEEEauthorblockA{\\Liverpool John Moores University, Byrom Street, Liverpool, L3 3AF, UK}
\IEEEcompsocitemizethanks{
	\IEEEcompsocthanksitem * Paul Fergus is with the Department of Computer Science, Liverpool John Moores University, Byrom Street, Liverpool, L3 3AF, UK. (E-mail: p.fergus@ljmu.ac.uk\protect\\
  \IEEEcompsocthanksitem Casimiro Curbelo Monta\~{n}ez is with the Department of Computer Science, Liverpool John Moores University, Byrom Street, Liverpool, L3 3AF, UK.\protect\\
  \IEEEcompsocthanksitem Basma Abdulaimma is with the Department Computer of Science, Liverpool John Moores University, Byrom Street, Liverpool, L3 3AF, UK.\protect\\
    \IEEEcompsocthanksitem Paulo Lisboa is with the Department of Mathematics, Liverpool John Moores University, Byrom Street, Liverpool, L3 3AF, UK.\protect\\
    \IEEEcompsocthanksitem Carl Chalmers is with the Department of Computer Science, Liverpool John Moores University, Byrom Street, Liverpool, L3 3AF, UK.\protect\\
}}

\markboth{}%
{Fergus \MakeLowercase{\textit{et al.}}: Bare Demo of IEEEtran.cls for Computer Society Journals}

\IEEEtitleabstractindextext{%
\begin{abstract}
Genome-Wide Association Studies (GWAS) are used to identify statistically significant genetic variants in case-control studies. The main objective is to find single nucleotide polymorphisms (SNPs) that influence a particular phenotype (i.e. disease trait). GWAS typically use a p-value threshold of $5*10^{-8}$ to identify highly ranked SNPs. While this approach has proven useful for detecting disease-susceptible SNPs, evidence has shown that many of these are, in fact, false positives. Consequently, there is some ambiguity about the most suitable threshold for claiming genome-wide significance. Many believe that using lower p-values will allow us to investigate the joint epistatic interactions between SNPs and provide better insights into phenotype expression. One example that uses this approach is multifactor dimensionality reduction (MDR), which identifies combinations of SNPs that interact to influence a particular outcome. However, computational complexity is increased exponentially as a function of higher-order combinations. This makes approaches like MDR difficult to implement. Even so, understanding epistatic interactions in complex diseases is a fundamental component for robust genotype-phenotype mapping. In this paper, we propose a novel framework, based on nonlinear transformations of combinatorically large SNP data, using stacked autoencoders, to identify higher-order SNP interactions. We focus on the challenging problem of classifying preterm births. Evidence suggests that this complex condition has a strong genetic component with unexplained heritability reportedly between 20\%-40\%. This claim is substantiated using a GWAS data set, obtained from dbGap, which contains predominantly urban low-income African-American women who had normal deliveries (between 37 and 42 weeks of gestation) and preterm deliveries (less than 37 weeks of gestation). Latent representations from original SNP sequences are used to initialize a deep learning classifier before it is fine-tuned for classification tasks (term and preterm births). The complete network models the epistatic effects of major and minor SNP perturbations. All models are evaluated using standard binary classifier performance metrics. The findings show that important information pertaining to SNPs and epistasis can be extracted from 4666 raw SNPs generated using logistic regression (p-value=$5*10^{-3}$) and used to fit a deep learning model and obtain results (Sen=0.9562, Spec=0.8780, Gini=0.9490, Logloss=0.5901, AUC=0.9745, MSE=0.2010) using 50 hidden nodes and (Sen=0.9289, Spec=0.9591, Gini=0.9651, Logloss=0.3080, AUC=0.9825, MSE=0.0942) using 500 hidden nodes.   
\end{abstract}

}

\maketitle

\IEEEdisplaynontitleabstractindextext

\IEEEpeerreviewmaketitle

\IEEEraisesectionheading{\section{Introduction}\label{sec:introduction}}

\IEEEPARstart {P}{reterm} birth (PTB) is the delivery of live babies born before 37 weeks of gestation \cite{Blencowe2013}. In contrast, term births are the live delivery of babies born between 37 and 42 weeks. In 2010, the World Health Organisation (WHO) declared that preterm deliveries accounted for 1 in 10 births worldwide \cite{Blencowe2013}. Compared with Caucasians, the risk of preterm birth in African-Americans is 1.5 times higher. This groups also has an even greater risk of giving birth before 32 weeks gestation \cite{Kistka2007}. Population-specific risk factors include anaemia during pregnancy, low serum folate levels, vitamin D deficiency, poor weight gain during pregnancy, and high pregnancy body mass index (BMI) \cite{Anum2009}. These are independent of socio-economic status or other social factors \cite{Goldenberg1996}, \cite{McGrady1992}.
\par
PTB has significant adverse effects on newborns. The severity increases the more premature the delivery is. Approximately, 50\% of all perinatal deaths are caused by preterm delivery. For those that survive, they often suffer with disorders caused by the birth, such as impairment to hearing, vision, the lungs and cardiovascular system, and non-communicable disease. Up to 40\% of survivors of extremely premature birth can also develop chronic lung disease \cite{Greenough2012}. In other cases, survivors suffer from neuro-developmental or behavioural defects, including cerebral palsy, motor, learning and cognitive impairments.   
\par
The precise etiology of PTB remains elusive. However, 30\%-35\% are known to be medically indicated (i.e. preeclampsia and foetal growth restriction) \cite{Goldenberg2008}. Preterm prelabor ruptured membranes (PPROMs - often attributed to infection, placental abruption, and anatomical abnormalities in the mother) account for 25\%-30\% \cite{Goldenberg2008} and spontaneous PTB (sPTB) for the remaining 35\%-45\% where the cause is unclear \cite{Moutquin2003}.      
\par
A strong body of evidence, from twin-based studies, has shown that maternal and foetal genetic factors contribute to PTB with heritability between 20\%-40\% \cite{Treloar2000}, \cite{Clausson2000}, \cite{Svensson2009}. Though attempts to identify the specific variant(s) of prematurity in genome-wide association studies (GWAS) have failed to produce any reproducible findings \cite{YORK2014}. Several GWAS studies have identified notable relationships but meta-analysis has shown that these are often negligible or contained within a particular population \cite{Hindorff2009}, \cite{DeFranco2007}.   
\par
Associations can be measured using Bonferroni correction, which is a highly conservative threshold designed to minimize type 1 errors in multiple testing studies. This leads to missing heritability were single genetic variations cannot fully explain the heritability of phenotypes \cite{Maher2008}, \cite{Wei2014}. Among the many approaches that exist, multifactor dimensionality reduction (MDR), however, has found that single nucleotide polymorphisms (SNPs) with little individual effect, through their interactions, can account for more variance in phenotypes \cite{Calle2010}, \cite{Wan2010}, \cite{Lishout2013}. Random forest algorithms have also been heavily utilised to detect significant SNPs in large-scale GWAS \cite{Bureau2005, Jiang2009, Schwarz2010, Yoshida2011, Kursa2014}. However, enumerating the large number of high-order combinations common in genetics is computationally very difficult to implement and a major limitation in these approaches. This issue has been mitigated by applying filters to select groups of SNPs that are relevant to the phenotype of interest \cite{Grady2012}, for example, using PLINK \cite{Purcell2007} which also provides two SNP epistatic analysis. Larger combinations are possible using LAMPLINK \cite{Terada2013}, but scalability issues still persist.  
\par
In this paper, we use a novel deep learning (DL) \cite{LeCun2015} framework to extract latent representations from original SNP sequences using a stacked autoencoder \cite{Ng2011}. A deep learning classifier is initialised using the generated stacked autoencoder model and fine-tuned to classify term and preterm observations. The complete network models the epistatic effects of major and minor SNP perturbations. Our approach demonstrates the potential of DL as a powerful framework for GWAS analysis that can capture information about SNPs and the important interactions between them for accurate inference on previously intractable models. This is the first comprehensive study of its kind that presents a framework consisting of stacked autoencoders and DL classification models in GWAS analysis. 
\par   
The remainder of this paper is organised as follows. Section 2 describes the Materials and Methods used in the study. The results are presented in Section 3 and discussed in Section 4 before the paper is concluded and future work is presented in Section 5.   

\section{Materials and Methods}
The data, for this study, was obtained through authorized access to dbGap (Study Accession: phs000332.v3.p2) \cite{Hao2004}. The data set includes 722 cases and 1057 controls. Cases were drawn from deliveries at the Boston Medical Center (BMC) that occurred before 37 weeks of gestation irrespective of birth weight. Controls include mothers who delivered term babies after 37 weeks of gestation also from the BMC cohort. Controls were frequency matched with case mothers on race, age ($\pm$ 5 years), and the baby's gender and parity. Women were excluded if pregnancies were due to vitro fertilization, they had multiple pregnancies, or the foetus had chromosomal abnormalities or major birth defects. Further exclusion criteria included mothers who had congenital or acquired uterus lesions, a known history of an incompetent cervix, or previous PTBs caused by maternal trauma. Each subject was interviewed using a standardized questionnaire to gather important epidemiological data, including ultrasound findings, placental pathology reports, laboratory reports, information on pregnancy complications and birth outcomes.

\subsection{Data Collection}
The GWAS recruited ~1000 mothers who delivered preterm and ~1000 age-matched mothers who had term births (African-American - 68\%; Haitian - 31.5\%). The subjects were genotyped in two phases. The first phase was completed in 2011 and the second in 2014. For all study samples, the Qiagen method was used to extract DNA from whole blood. In each phase cases and controls were balanced across 96-well plates and each plate contained between two and four HapMap controls, as well as an average of two study duplicates. Phase 1 was genotyped using the Illumina HumanOmni2.5-4v1 array and using the calling algorithm GenomeStudio version 2-10.2, Genotyping Module version 1.74 and GenTrain version 1.0. Phase 2 was genotyped using the Illumina HumanOmni2.5-8v1 array and using the calling algorithm GenomeStudio version 2011.1, Genotyping Module version 1.9.4 and GenTrain version 1.0. The two phases were merged into a single data set with 2,369,543 probes common to both arrays.   
\par
A total of 1,910 observations (including duplicates) from study subjects were put into genotype production, of which 1,889 were successfully genotyped and passed the Center for Inherited Disease Research (CIDR's) quality control (QC) process. The subsequent quality assurance (QA) procedure removed five observations, and the final set of scans posted to dbGAP included 1,884 study participants and 62 HapMap controls. The 1,884 study observations were derived from 1808 subjects and include 76 pairs of duplicate scans. The 62 HapMap control scans were derived from 24 subjects, all of which were replicated two or more times. The study subjects occur as 1,681 singletons and 60 families of 2-4 members each. The study families were discovered during the analysis of relatedness. The HapMap controls include 8 trios (4 CEU, 4 YRI).  

\subsection{Quality Control}
The data set was subjected to pre-established QC protocols as recommended in \cite{Anderson2010}, where data QC was applied to individuals first and then to the genetic variants. PLINK v1.9 \cite{Purcell2007} was used on a Linux Ubuntu machine, version 16.04 LTS, with 16GiB of Memory and an Intel Core I7-7500U CPU @ 2.70HHz x 4, to conduct the required QC and filtering procedures. Before QC, the 24 HapMap controls and the 0 Chromosome were removed from the data set.   
\par
Individual QC: Individuals with discordant sex information (homozygosity rate between 0.2 and 0.8) were identified using the X-chromosome and ascertained sex. This resulted in eight individuals being removed from the data set. Individuals with elevated missing data rates were identified using a genotype failure rate $\geq$ 0.02 (seven individuals were removed). While, outlying heterozygosity was identified using a heterozygosity rate $\pm$3 standard deviations from the mean (16 individuals were removed).Pairs of individuals with identity by descent (IBD) $>$ 0.185 were identified resulting in 38 individuals being removed. Principle Component Analysis (PCA) was conducted for the identification of outliers and hidden population structure using EIGENSOFT \cite{Price2006}. Individuals were identified with divergent ancestry using thresholds -0.05 and 0.00 for principal component (PC) 1 and 2 respectively. This resulted in 289 individuals being removed using the PC1 threshold and 297 using the PC2 threshold. All unique missing markers were combined and excluded from the data set reducing the total number of individuals to 1527 (Case=632, Control=895) with the genotyping rate in remaining samples equal to 0.992308.   
\par
Marker QC: Each individual contains 2,362,044 SNPs. SNPs with a significantly different ($p < 1*10^{-5}$) missing data rate between cases and controls were removed (n=22603) resulting in 2,339,441 remaining SNPs. SNPs with minor allele frequency (MAF $<1$\%), call rate $<$98\% and deviations from Hardy-Weinberg equilibrium ($p<1*10^{-5}$) were excluded. The data set following QC resulted in 1527 individuals with 1,927,820 variants each.   

\subsection{Association Analysis}
In this study, association analysis is used to reduce the computationally large number of SNPs (1,927,820) for machine learning tasks. Several p-value thresholds are considered that range between $5*10^{-3}$ and $5*10^{-8}$ inclusive - $5*10^{-8}$ being the Bonferroni correction \cite{dunn1959}. The resulting groups contain between 3 and 4666 SNPS (depending on the threshold) and are used to train and baseline classifier models. These models are then compared with a stacked autoencoder and deep learning framework with 4666 SNPs (obtained using $5*10^{-3}$) to force a neural network to extract latent representations from the SNPs. This transformation is designed to capture the epistatic interactions between SNPs that influence the phenotype (preterm birth).   

\subsubsection{Association Testing}
Using a standard association analysis procedure, \{X\textsubscript{1}, \dots, X\textsubscript{u}\} is a set of \textit{U} SNPs for \textit{N} individuals, and phenotypes are described as \{y\textsubscript{1}, \dots, {y\textsubscript{n}\}. In this study only one phenotype is considered (preterm birth). For each SNP, there is a minor allele \textit{a} and major allele \textit{A}. The homozygous major allele is defined as \textit{AA}, the heterozygous allele as \textit{Aa} and the homozygous minor allele as \textit{aa} - 0, 1, and 2 are used to describe these respectively. Therefore, X\textsubscript{un} $\in$ \{0,1,2\}, (1 $\leq$ \textit{u} $\leq$ \textit{U}, 1 $\leq$ \textit{n} $\leq$ \textit{N}). The phenotype is represented as a binary variable, 0 referring to controls and 1 referring to cases. 
\par
Genotypes are grouped into an additive model. Given $A$ we assume that there is a uniform, linear increase in risk for each copy of the $A$ allele. For example, if the risk is 3x for $Aa$ then the risk is 6x for $AA$. The additive model is only considered in this study as it has satisfactory power in detecting additive and dominant effects. 

\subsubsection{Logistic Regression}
Logistic regression is used to assess which SNPs increase the odds of a given outcome (in this study a preterm birth). This is performed under an additive model where logistic regression modelling for conditional probability \textit{Y} = 1 is: \cite{Wang2016}:   
\begin{equation} 
  \theta(X) = P(Y = 1|X)
\end{equation}
\par
The logit function which is the inverse of the sigmoidal logistic function, is represented as:
\begin{equation} 
  logit(X) = ln \frac{\theta(X)}{1 - \theta(X} 
\end{equation}
\par
The logit is given as a linear predictor function as follows:
\begin{equation} 
  logit(X)~\beta_{0} + \beta_{1}X
\end{equation}
Utilizing logistic regression, while not ideal, allows the number of SNPs with insignificant marginal effects to be reduced to meet the computational needs required for machine learning tasks. The remaining SNPs capture the linear interactions between SNPs and the phenotype but not the cumulative epistatic interactions that exist between the remaining SNPs. To capture epistatic interactions, we utilise a deep learning model pre-initialised with a stacked sparse autoencoder.   
\subsection{Deep Learning}
A multi-layer feedforward neural network is implemented in this study based on the formal definitions in \cite{Ng2011}. Computational units (neurons) take as input $(x_1, x_2, \dots x_n)$, and a +1 intercept term, and generate outputs $h_{W,b}(x) = f(W^Tx) = f(\sum^n_{i=1} W_ix_i+b)$ , where $f: \R \mapsto \R$ is the activation function. Under supervised learning conditions, uniform adaptive optimization governs weight initialization. A rectifier nonlinear activation function \textit{f} is implemented to control weight summing and node activation according to:   
\begin{equation} 
  \begin{aligned}
    f(x) = max(0,x),
\end{aligned}
\end{equation}
where $x$ is the input to a computational unit. 
\par
The network structure contains input, hidden and output layers where $n_l$ denote the number of layers and $L_l$ a particular layer. Parameters $(W,b)$ = $(W^{(1)}$, $b_{(1)}$, $W^{(n)})$, $b_{(n)})$ are described in the network where $W^{(l)}_{ij}$ denotes the weight of the synaptic connection between unit $j$ in layer $l$, and unit $i$ in layer $l+1$. An intercept node $b^{(l)}_i$, associated with unit $i$ in layer $l+1$ is introduced as a bias to overcome the problem associated with input patterns that are zero. The number of nodes in a layer is denoted by $s_l$ for $l$ (this does not include the bias unit). Thus, $a{(l)}_i$ refers to the activation of node $i$ in layer $l$. Given the parameters $W,b$, the neural network hypothesis is defined as $h_{W,b}(x)$ which outputs a real number.   
\par 
The network is trained using a sample set of observations $(x^{(i)}, y^{(i)})$ where $y^{(i)} \in \R^2$. With a fixed training set ${(x^{(1)}, y^{(1)}), \dots, (x^{(m)}, y^{(m)})}$ of $m$ examples, the neural network is trained using gradient descent and the cost function is calculated using:  

\begin{equation} 
  \begin{aligned}
    J(W,b) = \bigg[\frac{1}{m}\sum_{i=1}^{m}J(W,b,x^{(i)}, y^{(i)})\bigg] \\
           + \frac{\lambda}{2}\sum^{n_l-1}_{l=1}\sum^{s_l}_{i=1}\sum^{s_l+1}_{j=1}\big(W^{(l)}_{ji}\big)^2 \\ 
           = \bigg[\frac{1}{m}\sum_{i=1}^{m}\bigg(\frac{1}{2}||h_{w,b}(x^{(i)})-y^{(i)}||^2\bigg)\bigg]\\
           + \frac{\lambda}{2}\sum^{n_l-1}_{l=1}\sum^{s_l}_{i=1}\sum^{s_l+1}_{j=1}(W^{(l)}_{ji})^2
\end{aligned}
\end{equation}

where the first expression is the average sum of squared errors and the second, a weight decay term for decreasing the strength of weights, and  preventing overfitting. The relative importance of the two expressions is controlled with the weight decay parameter $\lambda$.  
\par
The parameters $W^{(l)}_{ij}$ and each $b^{(l)}_{i}$ are initialized to a random value close to zero before training commences. This is a necessary step that prevents hidden layer units learning the same function of the input. The cost function is used to minimize $J(W,b)$ and parameters $W,b$ are updated with:   

\begin{equation} 
  \begin{aligned}
    W^{(l)}_{ij} := W^{(l)}_{ij} - \alpha \frac{\partial}{\partial W^{(l)}_{ij}}J(W,b)\\
    b^{(l)}_i := b^{(l)}_i - \alpha \frac{\partial}{\partial b^{(l)}_{i}}J(W,b) 
\end{aligned}
\end{equation}

where $\alpha$ is the learning rate.  
\par
The backpropagation algorithm (see Algorithm 1) computes the partial derivatives $\frac{\partial}{\partial W^{(l)}{ij}} J(W,b;x,y)$ and $\frac{\partial}{\partial b^{(l)}{i}} J(W,b;x,y)$ of the cost function for a single sample $J(W,b;x,y)$. 

\begin{algorithm}[htp]
  \begin{algorithmic}[1]
    \caption{Backpropagation Algorithm}
      \STATE Perform forward pass and compute activations for $L_2$, \dots, $L_n$
      \FOR{i=1, \dots, $n_l$,}
      \STATE $\delta^{(n_l)}_i = \frac{\partial}{\partial^{(n_l)}_{z_i}} \frac{1}{2}||y - h_{W,b}(x)||^2 = - (y_i - a^{(n_l)}_i) \cdot f'(z^{(n_l)}_i)$
      \ENDFOR
      \FOR{l=$n_l-1$, \dots, 2,}
      \FOR{i=1, \dots, l}
      \STATE $\delta^{(l)}_i = \Big(\sum^{S_l+1}_{j=1} W^{(l)}_{ji}\delta^{(l+1)}_j\Big) f'(z^{(l)}_i)$
      \ENDFOR
      \ENDFOR
      \STATE Compute the desired partial derivatives:
      \STATE  $\frac{\partial}{\partial W^{(l)}_{ij}}J(W,b;x,y) = a^{(l)}_j \delta^{(l+1)}_i$
      \STATE  $\frac{\partial}{\partial b^{(l)}_{i}}J(W,b;x,y) = \delta^{(l+1)}_i$
  \end{algorithmic}
\end{algorithm}

Each node $i$ in layer $l$ is used to compute an error term $\delta^{(l)}_i$ and measure the node's contribution to errors that occurred in the output. With respect to output nodes, the error term $\delta^{(n_l)}_i$ (where layer $n_l$ is the output layer), represents the difference between the network's activation and the true target value. While hidden units compute a $\delta^{(l)}_i$ using a weighted average of the error terms of the nodes that use $a^{(l)}_i$ as input.  

Once the derivatives have been computed the derivatives for the overall cost function can be obtained using:

\begin{equation} 
  \begin{aligned}
    \frac{\partial}{\partial W^{(l)}{ij}} J(W,b) = \Big[\frac{1}{m} \sum^{m}_{i=1} \frac{\partial}{\partial W^{(l)}{ij}} J(W,b;x^{(i)},y{(i)}\Big]\\
    +\lambda W^{(l)}_{ij}\\
    \frac{\partial}{\partial b^{(l)}{i}} J(W,b) = \frac{1}{m} \sum^{m}_{i=1} \frac{\partial}{\partial W^{(l)}{i}} J(W,b;x^{(i)},y^{(i)})
\end{aligned}
\end{equation}

It is now possible to describe the gradient descent algorithm. In Algorithm 2 $\Delta W^{(l)}$ is a matrix with a dimension equal to $W^{(l)}$, and $\Delta b^{(l)}$ is a vector with a dimension equal to $b^{(l)}$.

\begin{algorithm}[htp]
  \begin{algorithmic}[1]
    \caption{Gradient Decent}
      \STATE Set $\Delta W^{(l)} := 0$, $\Delta b^{(l)} := 0$ (matrix/vector of zeros) for all $l$.
      \FOR{i=1, \dots, m,}
      \STATE Use backpropagation to compute $\bigtriangledown _{W^{(l)}}J(W,b;x,y)$ and $\bigtriangledown_{b^{(l)}}J(W,b;x,y)$ 
      \STATE Set $\Delta W^{(l)} := \Delta W^{(l)} + \bigtriangledown_W^{(l)}J(W,b;x,y).$
      \STATE Set $\Delta b^{(l)} := \Delta b^{(l)} + \bigtriangledown_b^{(l)}J(W,b;x,y).$
      \ENDFOR
      \STATE Update the parameters:
      \STATE $W^{(l)} := W^{(l)} - \alpha \Big[\Big(\frac{1}{m}\Delta W^{(l)}\Big) + \lambda W^{(l)}\Big]$
      \STATE $b^{(l)} := b^{(l)} - \alpha \Big[\frac{1}{m} \Delta b^{(l)}\Big]$
  \end{algorithmic}
\end{algorithm}

The neural networks in this study are trained for classification tasks with repeated steps of gradient descent to reduce the cost function $J(W,b)$. 

\subsection{Stacked Sparse Autoencoders}

A stacked sparse autoencoder (SSAE), based on the previously defined deep learning formalisms is implemented to further reduce the dimensionality of the subset of SNPs generated using logistic regression (p-value threshold ($5*10^{-3}$ - 4666 SNPs). The primary goal is to extract the latent information from the 4666 SNPs and produce a significantly smaller input feature space for classifier modelling and case-control classification tasks. The optimal hidden layer units are utilised to achieve this, such that the output $\hat{x}$ is similar to the input $x$:   

\begin{equation} 
  \begin{aligned}
      h_{W,b}(x) \approx x
  \end{aligned}
\end{equation}

The hidden unit activations $a^{2}$ in $\R^{h}$ aim to reconstruct the input $x$. If there is structure in the data, the autoencoder will learn it. In fact, very simple autoencoders often learn a low-dimensional representation similar to principal component analysis (PCA). 
\par
Nodes in the autoencoder fire when output values are close to 1 and remain inactive when the output is close to 0. The goal is to ensure that nodes remain mostly inactive. Thus, the activation of a hidden unit is represented as $a^{(2)}_{j}(x)$ when the network receives input $x$. We let:   

\begin{equation} 
  \begin{aligned}
     \hat{p}_j = \frac{1}{m} \sum^{m}_{i=1} \Big[a^{(2)}_{j}(x^{(i)})\Big] 
  \end{aligned}
\end{equation}

represent the average activation of hidden unit $j$ and enforce the constraint:

\begin{equation} 
  \begin{aligned}
     \hat{p}_{j} = p 
  \end{aligned}
\end{equation}

where $p$ is typically a small sparsity parameter close to zero (for example, $p = 0.05$). In order to meet this constraint, the activation of the hidden unit should be mostly 0. To achieve this a penalty term is added that penalizes $\hat{p}_{j}$ when it deviates significantly from $p$:

\begin{equation} 
  \begin{aligned}
     \sum^{s_2}_{j=1} p \text{ } log \frac{p}{\hat{p}_j} + (1 - p) log \frac{1-p}{1-\hat{p}_j} 
  \end{aligned}
\end{equation}

where $s_{2}$ is the number of units in the hidden layer, and $j$ an index used to sum the hidden units in the network. Kullback-Leibler (KL) divergence is used to enforce this penalty term:

\begin{equation} 
  \begin{aligned}
     \sum^{s_2}_{j=1} KL(p||\hat{p}_{j}) 
  \end{aligned}
\end{equation}

where $KL(p||p_{j})$ = $\frac{p}{\hat{p}_j} + (1 - p) log \frac{1-p}{1-\hat{p}_j}$ is the KL-divergence between two Bernoulli random variables with mean $p$ and $\hat{p}_{j}$. In this way KL-divergence is used to measure the difference between two distributions. This penalty function is either $KL(p||\hat{p}_{j})$ = 0 if $\hat{p} = p$, or it increases monotonically as $\hat{p}_{j}$ diverges from $p$. The cost function can now be defined as:

\begin{equation} 
  \begin{aligned}
     J_{sparse}(W,b) = J(W,b) + \beta \sum^{s_2}_{j=1} KL(p||\hat{p}_{j})
  \end{aligned}
\end{equation}

where $J(W,b)$ is the same as we previously defined and $\beta$ is used to control the weight of the sparsity penalty term. The term $\hat{p}_j$ is dependent on $W,b$, as it is the average activation of hidden unit $j$, and hidden unit activations are dependent on the parameters $W,b$. 
\par 
The KL-divergence term is incorporated into the previously defined derivative calculation and now computed as:   

\begin{equation} 
  \begin{aligned}
     \delta^{(2)}_i = \Big( \Big(\sum^{s_2}_{j=1} W^{(2)}_{ji} \delta^{(3)}_j \Big) + \beta \Big(-\frac{p}{\hat{p}_i} + \frac{1-p}{1-\hat{p}_i}\Big) \Big) f'(z^{(2)}_i)
  \end{aligned}
\end{equation}

It is important to know $\hat{p}_i$ to compute this term. After computing $\hat{p}_i$, a forward pass on each example is performed to allow backpropagation on that example. Therefore, you compute a forward pass twice on each example in your training set, which does make it computationally less efficient. 
\par
A single autoencoder is simple, due to its shallow structure. Consequently, a single-layer autoencoder's representational power is very limited. In this study, autoencoders are stacked to enable greedy layer wise learning where the $i_{th}$ hidden layer is used as input to the i+1 hidden layer in the stack. The results produced by the stacked autoencoder are utilized to pretrain (initialize) the weights for our deep learning network to classify term and preterm deliveries. The GWAS quality control, Logistic regression, deep learning classifier and the stacked autoencoder form the constituent components within our proposed framework.    

\subsubsection{Performance Measures}

Sensitivity and specificity are used in this study to represent the number of correctly identify case and control instances. Sensitivity describes the true positive rate (Controls - term deliveries) and Specificity the true negative rate (Cases - preterm deliveries).
\par
The area under the curve (AUC) is the probability that, for each pair of examples, one for each class, the example from the positive class will be ranked higher. If a classifier provides estimates according to $p(C_{i}|x)$, it is possible to obtain values $\{a1, \dots, a_{n}1, b1, \dots, b_{n}1\}$ and $\{b_{1}, \dots, b, \dots, b_{n2}; b_{i} = p(C_{2}|x), x_{i} \in C_{2}\}$ and measure how well separated the distributions $\hat{p}(x)$ for class $C_{1}$ and $C_{2}$ patterns are.   
\par
Using the estimates $\{a1, \dots, a_{n}1, b1, \dots, b_{n}1\}$ they can be ranked in increasing order. The class $C_{1}$ test points can be summed to see that the number of pairs of points, one from class $C_{1}$ and one from $C_{2}$ with $\hat{p}(x)$ smaller for class $C_{2}$ than the $\hat{p}(x)$ value for class $C_{1}$ is:

\begin{equation} 
    \sum_{i=1}^{n_{1}}(r_{i} - i) = \sum_{i=1}^{n_{1}}r_{i} - \sum_{i=1}^{n_{1}}i= S_{0} - \frac{1}{2}n_{1}(n_{1}+1)
\end{equation}

where $r_{i}$ is the ranked estimate, $S_{0}$ is the sum of the ranks of the class $C_{1}$ test patterns. Since there are $n_{1}n_{2}$ pairs, the estimate of the probability that a randomly chosen class $C_{2}$ pattern has a lower estimated probability of belonging to class $C_{1}$ than a randomly chosen class $C_{1}$ is:

\begin{equation} 
    \hat{A} = \frac{1}{n_{1}n_{2}}\left(S_{0} - \frac{1}{2}n_{1}(n_{1} + 1)\right)
\end{equation}

This is equivalent to the area under the ROC which provides an estimate obtained using the rankings alone and not threshold values to calculate it \cite{Web2011}.
\par
The Gini coefficient is often used in binary classification studies and is closely related to the AUC. The Gini coefficient is defined as being the area between the diagonal and the ROC curve:

\begin{equation} 
  \begin{aligned}
     Gini + 1 = 2 + AUC
  \end{aligned}
\end{equation}

The Gini coefficient measures statistical dispersion. The SNP(s) with a Gini coefficient of 1, predicts the data perfectly. A coefficient of 0 indicates that the SNP(s) have no predictive capacity. 
\par
Log Loss provides a measure of accuracy for a classifier whereby penalties are imposed on classifications that are false. Minimising the Log Loss is correlated with accuracy (as one increases the other decreases). Log loss is calculated by assigning a probability to each class rather than stating what the most likely class would be:
\begin{equation} 
  \begin{aligned}
     logloss = -\frac{1}{N} \sum^N_{i=1}[y_i log (p_i) + (1-y_i)log(1-pi)].
  \end{aligned}
\end{equation}

where $N$ is the number of samples, $y_{i}$ is a binary indicator for whether $j$ correctly classifies instance $i$. For models that classify all instances correctly the Log Loss value with be zero. For misclassifications, the Log Loss value will be progressively larger. 
\par
The Mean Squared Error (MSE) metric is utilised to measure the average sum of the square difference between actual values and predicted values for all data points. A MSE value of 0 indicates that the model correctly classifies all class instances. Again, for misclassifications, the MSE will be progressively larger.

\section{Results}
We first present the results using a deep learning classification model comprising four hidden layers with 10 neurons in each to provide baseline results. Several association analysis p-value filters are considered for dimensionality reduction - resulting SNP combinations are used to fit our classifier models. The performance of each model is measured using Sensitivity, Specificity, Gini, AUC, LogLoss and MSE values. The data set is split randomly into training (80\%), validation (10\%) and testing (10\%).   
\par
A RectifierWithDropout activation function is used over one hundred epochs. The learning rate is configured to 0.005 with rate annealing set to $1*10^{-6}$ and rate decay set to 1. The learning rate is a function of the difference between the predicted value and the target value. This is a delta value available at the output layer. The output at each hidden layer is corrected using backpropagation. Rate annealing is utilised during this process to reduce the learning rate to freeze into local minima. Rate decay simply controls the change of learning rate across layers. Momentum start is set to 0.5 and momentum ramp and momentum stable to $1*10^{-6}$ and 0 respectively. Momentum start controls the amount of momentum at the beginning of training. While momentum ramp controls the amount of learning for which momentum increases. Momentum stable simply controls the final momentum value reached after momentum ramp training examples.       

\subsection{Baseline Deep Learning Network}
\subsubsection{Classifier Performance}

Table 1 provides the performance metrics for the validation set. Metric values for association analysis p-values $5*10^{-3}$, $5*10^{-4}$, $5*10^{-5}$, $5*10^{-6}$, $5*10^{-7}$, and $5*10^{-8}$) were obtained using optimized F1 threshold values 0.6840 (resulting in 4666 SNPs), 0.6039435 (419 SNPs), 0.2799 (51 SNPs), 0.4471 (11 SNPs), 0.4107814 (11 SNPs) and 0.4450064 (three SNPs), respectively.  

\begin{table}[htp]
\renewcommand{\arraystretch}{1.3}
\caption{PERFORMANCE FOR VALIDATION SET}
\label{PMV}
\centering
\begin{tabular}{ccccccc}
      \hline\hline
      \bfseries p-value & \bfseries Sens & \bfseries Spec & \bfseries Gini & \bfseries LogLoss & \bfseries AUC & \bfseries MSE\\
      \hline\hline
      $5*10^{-3}$ & 0.9848 & 1.0000 & 0.9993 & 0.1002 & 0.9996 & 0.0150\\
      $5*10^{-4}$ & 0.9696 & 0.9285 & 0.9700 & 0.2988 & 0.9850 & 0.0597\\
      $5*10^{-5}$ & 0.7121 & 0.7959 & 0.6020 & 0.5679 & 0.8010 & 0.1928\\
      $5*10^{-6}$ & 0.9242 & 0.3673 & 0.3766 & 0.6669 & 0.6883 & 0.2369\\
      $5*10^{-7}$ & 0.9393 & 0.3265 & 0.3722 & 0.6507 & 0.6861 & 0.2290\\
      $5*10^{-8}$ & 0.8484 & 0.2959 & 0.2719 & 0.6745 & 0.6359 & 0.2407\\
      \hline\hline
\end{tabular}
\end{table}

Table 2 shows the performance metrics obtained using the trained models and the test data. The network comprised four hidden layers each containing 10 nodes. Based on empirical analysis this configuration produced the best results. Metric values for association analysis p-values $5*10^{-3}$, $5*10^{-4}$, $5*10^{-5}$, $5*10^{-6}$, $5*10^{-7}$, and $5*10^{-8}$ were again obtained using an optimized F1 threshold with values 0.7350, 0.3144, 0.2799, 0.4546975, 0.42307, and 0.4534978 respectively. The results are lower than those obtained by the validation set but in some cases not by much.   

\begin{table}[htp]
\renewcommand{\arraystretch}{1.3}
\caption{PERFORMANCE FOR TEST SET}

\label{PMT}
\centering
\begin{tabular}{ccccccc}
      \hline\hline
      \bfseries p-value & \bfseries Sens & \bfseries Spec & \bfseries Gini & \bfseries LogLoss & \bfseries AUC & \bfseries MSE\\
      \hline\hline
      
      $5*10{-3}$ & 1.0000 & 0.9882 & 0.9996 & 0.0960 & 0.9998 & 0.0128\\
      $5*10{-4}$ & 0.9000 & 0.9411 & 0.9388 & 0.3038 & 0.9694 & 0.0673\\
      $5*10{-5}$ & 0.9666 & 0.5411 & 0.6709 & 0.5581 & 0.8354 & 0.1913\\
      $5*10{-6}$ & 0.9333 & 0.3673 & 0.3766 & 0.6669 & 0.6883 & 0.2369\\
      $5*10{-7}$ & 0.9166 & 0.4352 & 0.4833 & 0.6374 & 0.7416 & 0.2225\\
      $5*10{-8}$ & 0.8833 & 0.4117 & 0.3572 & 0.6679 & 0.6786 & 0.2374\\
      \hline\hline
\end{tabular}
\end{table}

Early stopping (when misclassification rate converges) is adopted to avoid overfitting using the validation set as shown in Figure 1 (p-values $5*10^{-7}$ and $5*10^{-8}$ were omitted as the figure demonstrates a sharp deterioration in performance as the p-value is increased). Epochs represent the inflection points where performance on the validation set starts to decrease while performance on the training set continues to improve as the model starts to overfit. An optimised loss function is adopted to train the models. The AUC plots provide useful information about early divergence between the training and validation curves and highlight if overfitting occurs. From Figure 1, clearly there is a small amount of overfitting but nothing in excess.   
\par

\subsubsection{Model Selection}
The ROC curve in Figure 2 shows the cutoff values for the false and true positive rates using the test set. In this first evaluation, Figure 2 shows a clear deterioration in performance as the p-value filter is increased. Note that conservative p-value thresholds and the resulting SNPs, are an indication of how significant associations are. In this instance, machine learning demonstrates that highly ranked SNPs do not have sufficient predictive capacity to make distinctions between case and control instances (mothers who had term and preterm deliveries).  

\begin{figure}[htp] 
    \centering
  \subfloat[Logloss $5*10^{-3}$]{%
       \includegraphics[width=0.44\linewidth]{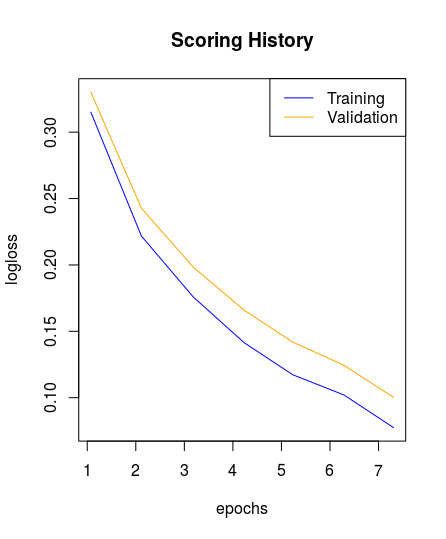}}
  \subfloat[AUC $5*10^{-3}$]{%
        \includegraphics[width=0.44\linewidth]{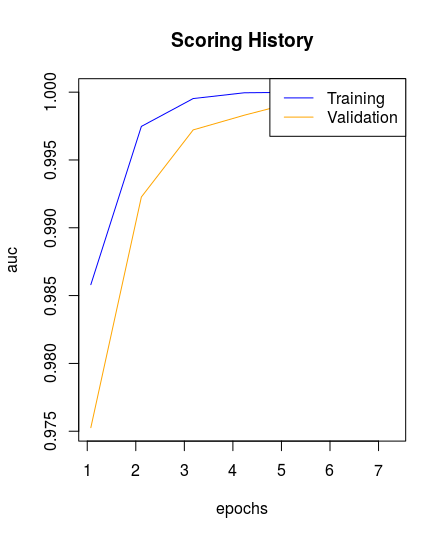}}\\
  \subfloat[Logloss $5*10^{-4}$]{%
        \includegraphics[width=0.44\linewidth]{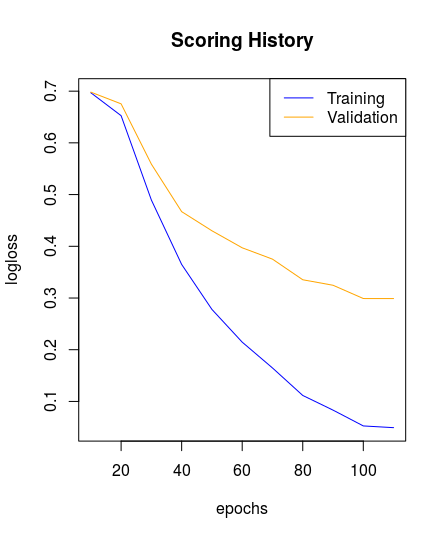}}
  \subfloat[AUC $5*10^{-4}$]{%
        \includegraphics[width=0.44\linewidth]{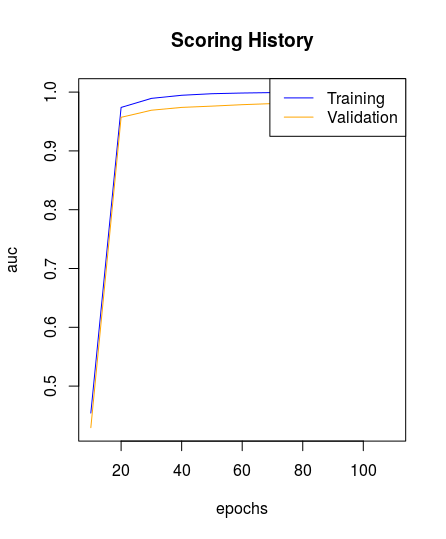}}\\
  \subfloat[Logloss $5*10^{-5}$]{%
        \includegraphics[width=0.44\linewidth]{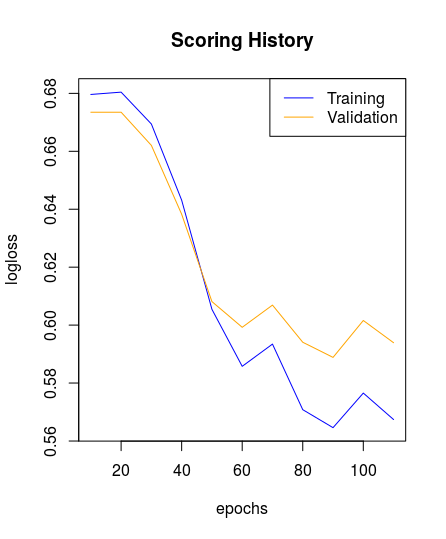}}
  \subfloat[AUC $5*10^{-5}$]{%
        \includegraphics[width=0.44\linewidth]{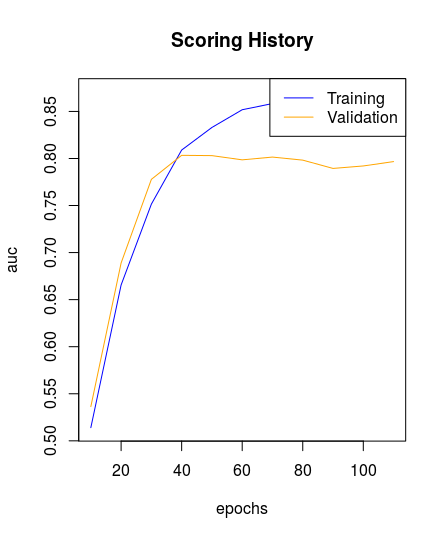}}\\
  \subfloat[Logloss $5*10^{-6}$]{%
        \includegraphics[width=0.44\linewidth]{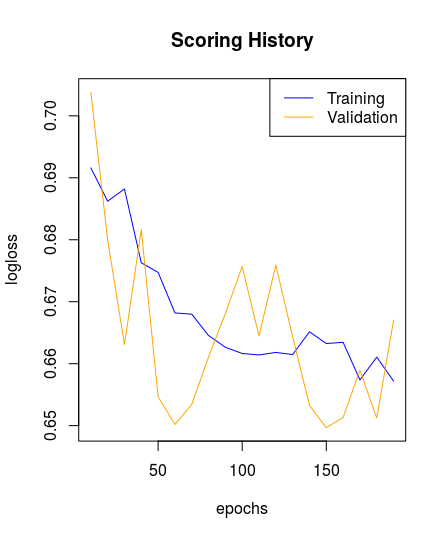}}
  \subfloat[AUC $5*10^{-6}$]{%
        \includegraphics[width=0.44\linewidth]{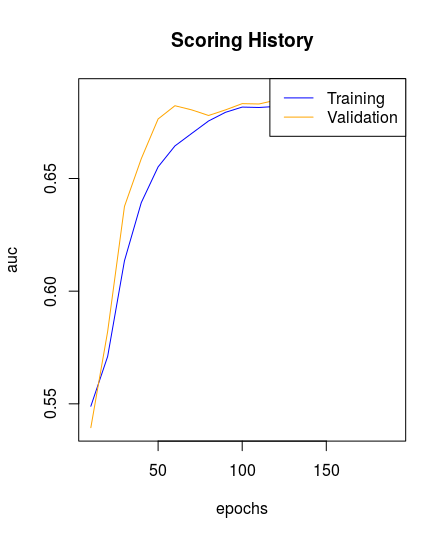}}\\
  \caption{(a) to (h) Logloss and AUC plots against epochs for p-value $5*10^{-3}$ to $5*10^{-6}$.}
  \label{fig1} 
\end{figure}

\begin{figure}[htp] 
    \centering
  \subfloat[ROC for $5*10^{-3}$]{%
       \includegraphics[width=0.44\linewidth]{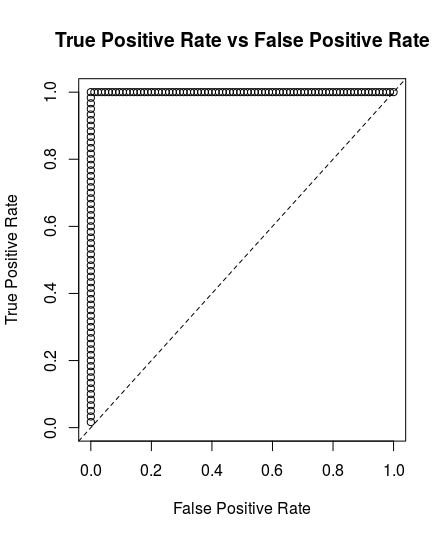}}
  \subfloat[ROC for $5*10^{-4}$]{%
        \includegraphics[width=0.44\linewidth]{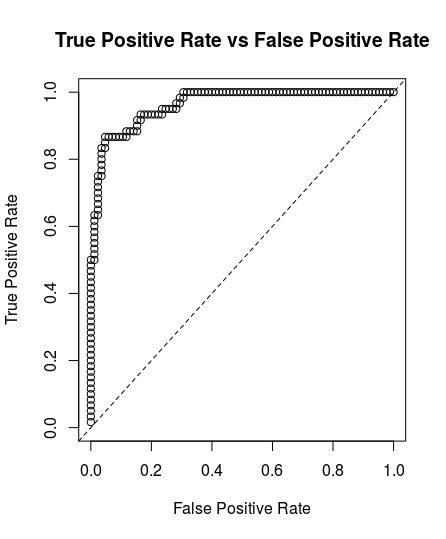}}\\
  \subfloat[ROC for $5*10^{-5}$]{%
        \includegraphics[width=0.44\linewidth]{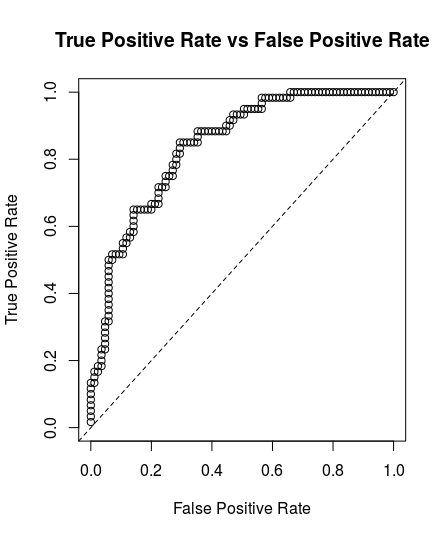}}
  \subfloat[ROC for $5*10^{-6}$]{%
        \includegraphics[width=0.44\linewidth]{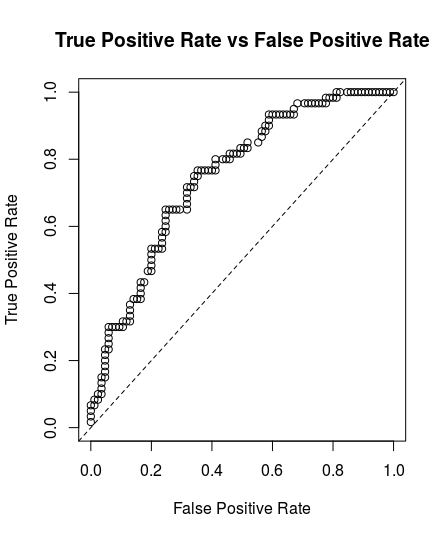}}\\
  \subfloat[ROC for $5*10^{-7}$]{%
        \includegraphics[width=0.44\linewidth]{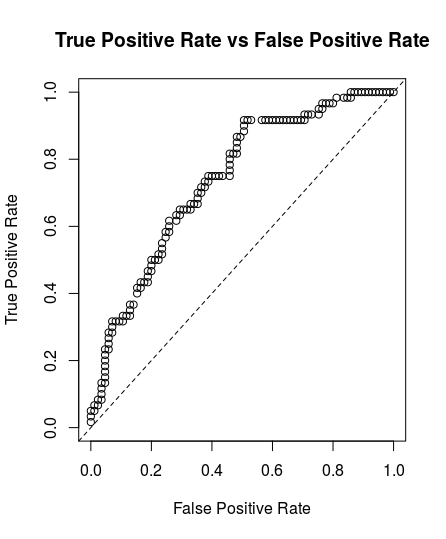}}
  \subfloat[ROC for $5*10^{-8}$]{%
        \includegraphics[width=0.44\linewidth]{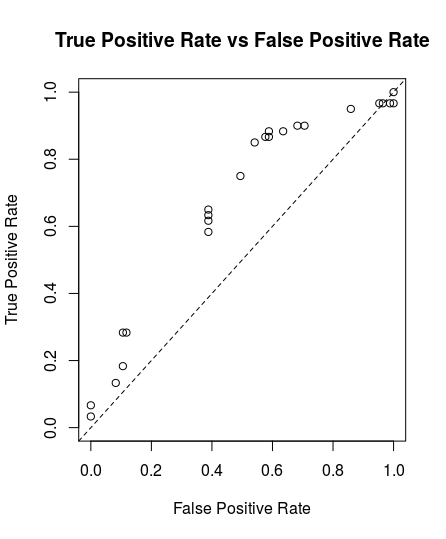}}\\
  \caption{(a) to (f) Performance ROC curves for test set using trained models and p-value between $5*10^{-7}$ and $5*10^{-8}$}.
  \label{fig2} 
\end{figure}

\subsection{Stacked Sparse Autoencoder}
In comparison, the following evaluation uses SNPs generated with a p-value $5*10^{-3}$. Latent features are extracted from 4666 SNPs with a stacked autoencoder that captures information about important SNPs and the cumulative epistatic interactions between them. This is achieved layer-wise by stacking simpler autoencoders that each contain a single hidden layer with 2000, 1000, 500, 200, 100 and 50 hidden nodes respectively. Deep learning classifiers are initialized with each of these layers and fine-tuned to classify case-control instances in the validation and test sets using four hidden layers with 10 nodes each.   
\subsubsection{Classifier Performance}
With the first layer (2000 neurons) a deep learning classifier model is initialised and then fine-tuned. The learning rate is set to $1*10^{-3}$ and an optimized F1 value of 0.7374 is used to extract metric values for the validation set as shown in Table 3. Subsequent layers are used to initialise and fine tune the remaining models with 1000, 500, 200, 100 and 50 hidden layers respectively. Metrics were obtained from these models using optimised F1 values 0.2979, 0.0769, 0.5881, 0.4996, and 0.6178 respectively. The learning rate for each of the layers is set to $1*10^{-3}$, $1*10^{-4}$, $1*10^{-5}$, $1*10^{-5}$, and $1*10^{-6}$. The full results are shown in Table 3   
\begin{table}[htp]
\renewcommand{\arraystretch}{1.3}
\caption{Performance Metrics for Validation Set}
\label{table_example}
\centering
\begin{tabular}{ccccccc}
      \hline\hline
      \bfseries Comp & \bfseries Sens & \bfseries Spec & \bfseries Gini & \bfseries LogLoss & \bfseries AUC & \bfseries MSE\\
      \hline\hline
      2000 & 0.9482 & 0.9772 & 0.9764 & 0.1273 & 0.9882 & 0.0331\\ 
      1000 & 0.9827 & 0.9659 & 0.9698 & 0.1246 & 0.9849 & 0.0270\\ 
      500 & 0.9827 & 0.8750 & 0.9674 & 0.3059 & 0.9837 & 0.0962\\ 
      200 & 0.8965 & 0.9545 & 0.9365 & 0.3752 & 0.9682 & 0.1098\\ 
      100 & 0.9482 & 0.7840 & 0.8475 & 0.6059 & 0.9237 & 0.2068\\ 
      50 & 0.9655 & 0.9545 & 0.9518 & 0.5854 & 0.9759 & 0.1988\\ 
      \hline\hline
\end{tabular}
\end{table}

Table 4 shows the performance metrics obtained using the test set. Layers containing 2000, 1000, 500, 200, 100, and 50 were again used with optimized F1 values 0.5836, 0.1695, 0.2348, 0.6036, 0.5061, and 0.5457 respectively. The learning rate values from the validation set were retained. The results are lower than those achieved with the validation set but in some cases not by much.   

\begin{table}[htp]
\renewcommand{\arraystretch}{1.3}
\caption{Performance Metrics for Test Set}
\label{table_example}
\centering
\begin{tabular}{ccccccc}
      \hline\hline
      \bfseries Comp & \bfseries Sens & \bfseries Spec & \bfseries Gini & \bfseries LogLoss & \bfseries AUC & \bfseries MSE\\
      \hline\hline
      2000 & 0.9672 & 0.9771 & 0.9939 & 0.0850 & 0.9969 & 0.0226\\ 
      1000 & 0.9781 & 0.9505 & 0.9736 & 0.1352 & 0.9868 & 0.0335\\ 
      500 & 0.9289 & 0.9581 & 0.9651 & 0.3080 & 0.9825 & 0.0942\\ 
      200 & 0.8797 & 0.8897 & 0.9055 & 0.3920 & 0.9527 & 0.1178\\ 
      100 & 0.8907 & 0.8593 & 0.8544 & 0.5970 & 0.9272 & 0.2024\\ 
      50 & 0.9562 & 0.878 & 0.9490 & 0.5901 & 0.9745 & 0.2010\\ 
      \hline\hline
\end{tabular}
\end{table}

Early stopping was again adopted to avoid overfitting as shown in Figure 3 (for brevity only layers 2000 to 200 are illustrated to show overfitting is appropriately managed). Again, there is a small amount of overfitting but nothing significant.

\subsubsection{Model Selection}
This time the ROC curve in Figure 4 shows significant improvements using the latent information captured in the hidden nodes. There is obvious deterioration, however, the results still remain high at 50 and only slightly worse that the results produced when 2000 hidden units are used.
\begin{figure}[htp] 
    \centering
  \subfloat[Logloss for hidden=2000]{%
       \includegraphics[width=0.44\linewidth]{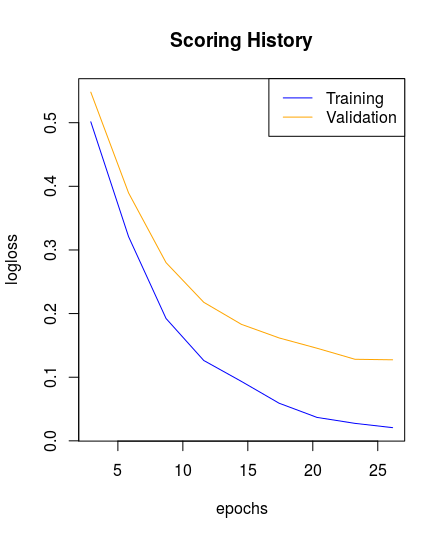}}
  \subfloat[AUC for hidden=2000]{%
        \includegraphics[width=0.44\linewidth]{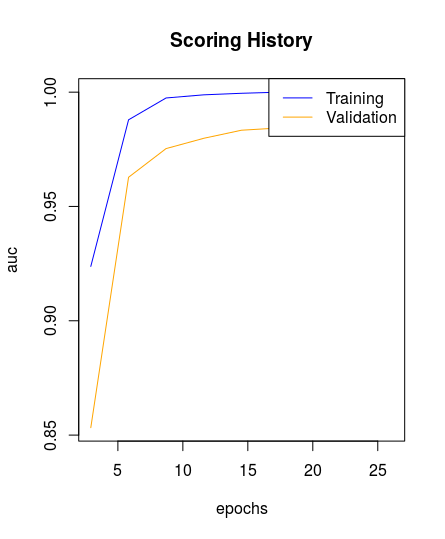}}\\
  \subfloat[Logloss for hidden=1000]{%
        \includegraphics[width=0.44\linewidth]{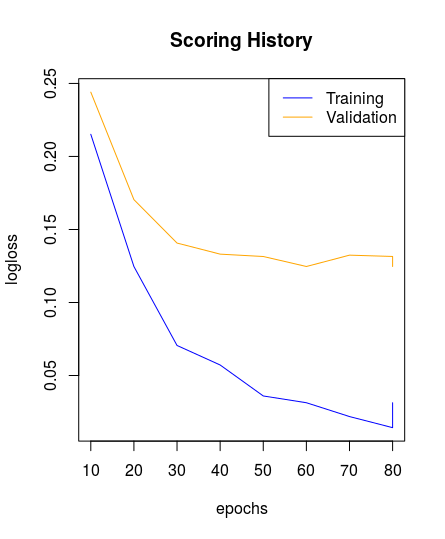}}
  \subfloat[AUC for hidden=1000]{%
        \includegraphics[width=0.44\linewidth]{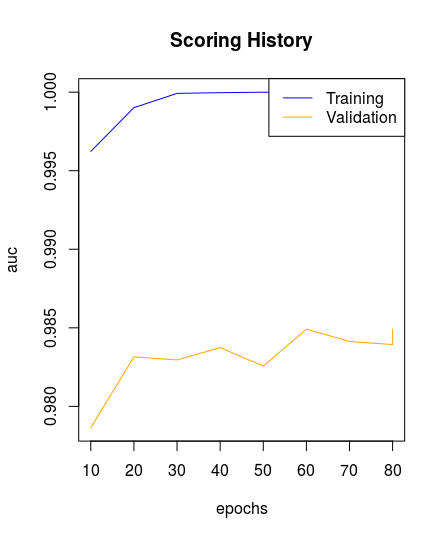}}\\
  \subfloat[Logloss for hidden=500]{%
        \includegraphics[width=0.44\linewidth]{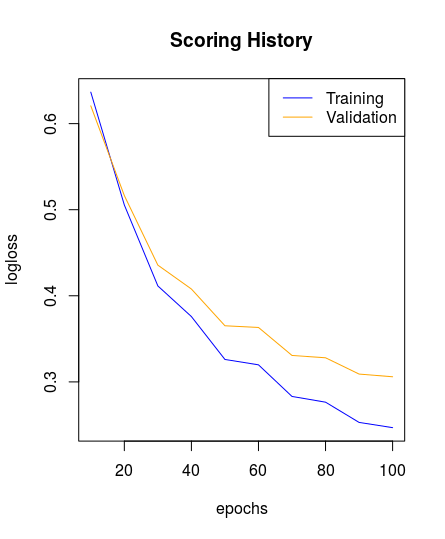}}
  \subfloat[AUC for hidden=500]{%
        \includegraphics[width=0.44\linewidth]{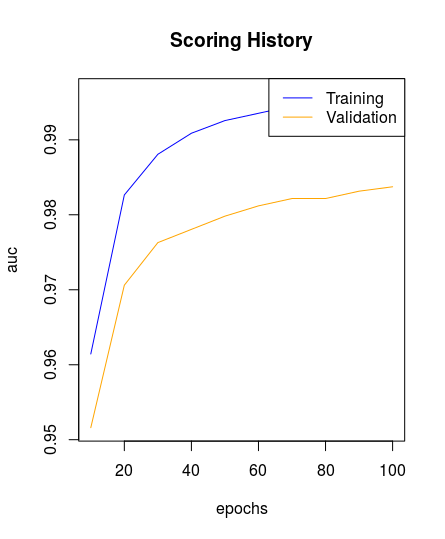}}\\
  \subfloat[Logloss for hidden=200]{%
        \includegraphics[width=0.44\linewidth]{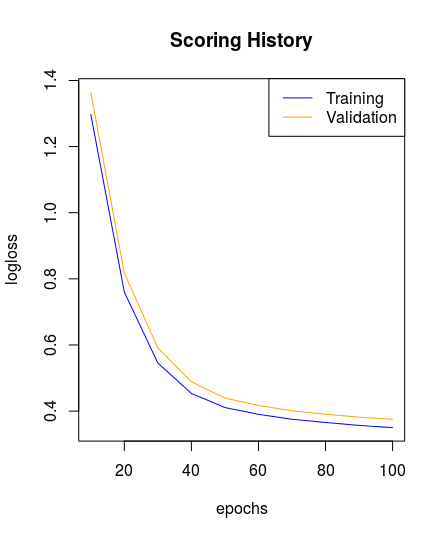}}
  \subfloat[AUC for hidden=200]{%
        \includegraphics[width=0.44\linewidth]{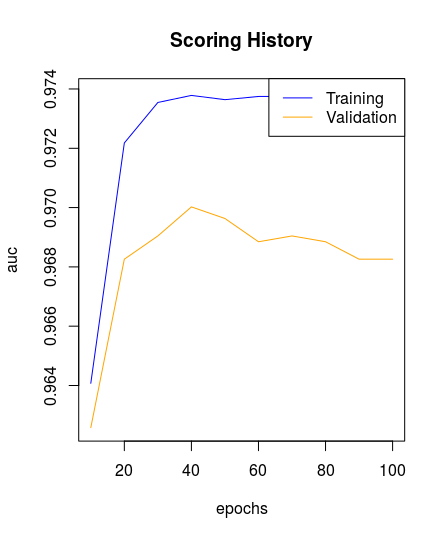}}\\
  \caption{(a) to (h) Logloss and AUC plots against epochs for 2000 to 200 Compression.}
  \label{fig3} 
\end{figure}

\section{Discussion}
In this paper, we have attempted to show the potential of deep learning as a novel framework for preterm birth GWAS analysis. The findings are encouraging. One important advantage deep learning has is its ability to abstract large, complex and unstructured data into latent representations that captures important information about SNPs and the epistatic interactions between them. This offers a powerful way to analyse GWAS data. Feature extraction is performed as a single unified process using a stacked autoencoder where multiple layers capture nonlinear dependencies and epigenetic interactions. These features do not differ when presented with small input changes. Consequently, this has the effect of eliminating noise and increasing robustness within the feature extraction process.   
\begin{figure}[htp] 
    \centering
  \subfloat[ROC for hidden=2000]{%
       \includegraphics[width=0.44\linewidth]{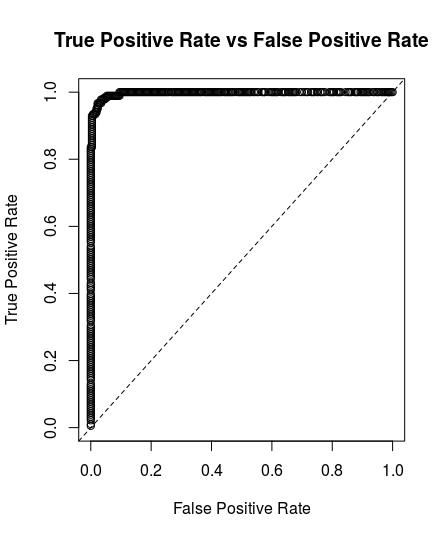}}
  \subfloat[ROC for hidden=1000]{%
        \includegraphics[width=0.44\linewidth]{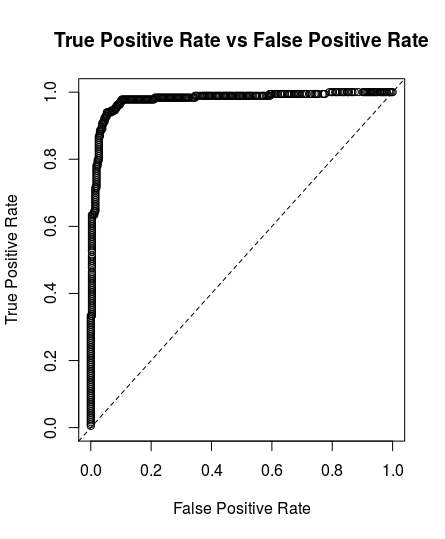}}\\
  \subfloat[ROC for hidden=500]{%
        \includegraphics[width=0.44\linewidth]{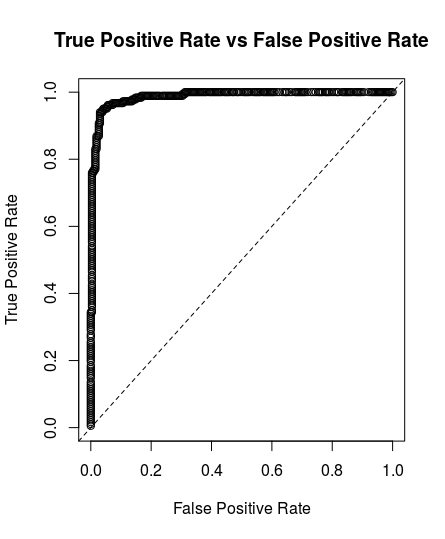}}
  \subfloat[ROC for hidden=200]{%
        \includegraphics[width=0.44\linewidth]{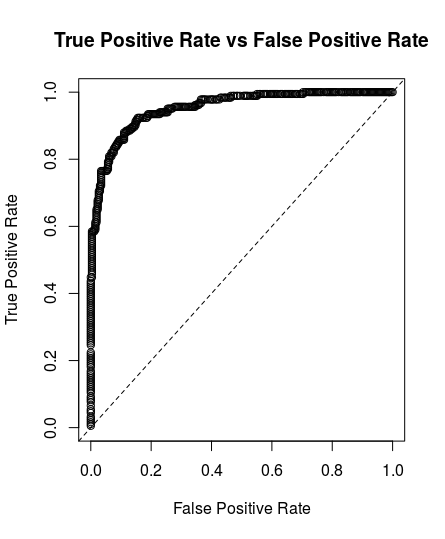}}\\
  \subfloat[ROC for hidden=100]{%
        \includegraphics[width=0.44\linewidth]{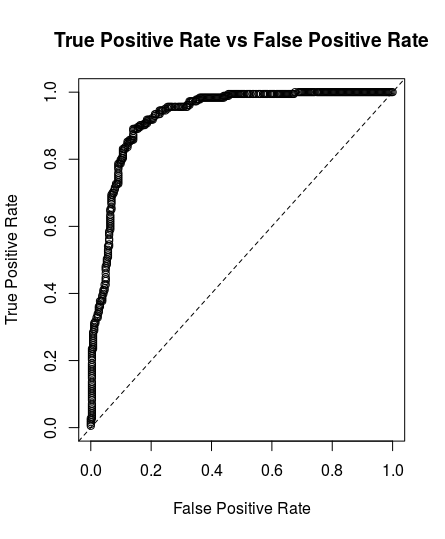}}
  \subfloat[ROC for hidden=50]{%
        \includegraphics[width=0.44\linewidth]{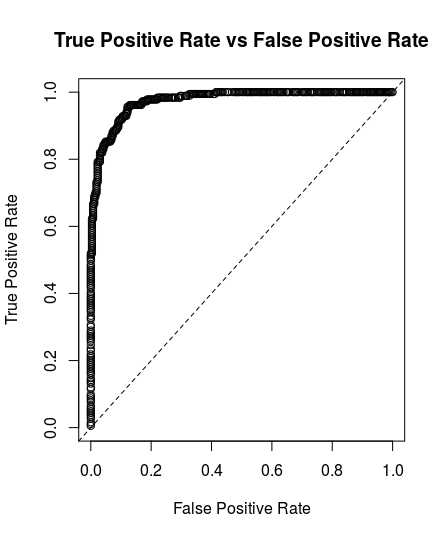}}\\
  \caption{(a) to (f) Performance ROC curves for test set using trained models.}
  \label{fig4} 
\end{figure}
\par
While, GWAS is useful for locating common variants of small effect and identifying very rare variants of much larger effect, they fail to classify phenotypes using suggestive or Bonferroni significance genome-wide associations. This is primarily caused by the fact that highly ranked SNPs are often false positives. Therefore, it is generally agreed that it may be possible to increase the proportion of variation captured in GWAS by incorporating information from rarer SNPs. Methods, such as MDR have attempted this, but they have been plagued by computational challenges.   
\par
Using a deep learning classifier model and a p-value threshold of $5*10^{-3}$ (4666 SNPs) it was possible to obtain good results (Sens=1, Spec=0.9882, Gini=0.9996, Log Loss=0.0960, AUC=0.9998, MSE=0.0128). However when the Bonferroni threshold ($5*10^{-8}$) is used, the results significantly drop (Sens=0.8833, Spec=0.4117, Gini=0.3572, Log Loss=0.0.6679, AUC=0.6786, MSE=0.2374). Clearly analysing single loci and their effect on a phenotype fails to capture the accumulative effects of less significant SNPs and their contribution to the outcome.   
\par
Therefore, a stacked autoencoder was utilised to extract the latent information from the 4666 SNPs through progressively smaller layers (2000, 1000, 500, 200, 100 and 50). The results using the test set showed significant improvement in classification accuracies. The best result was achieved using 2000 features (Sens=0.9672, Spec=0.9771, Gini=0.9939, Log Loss=0.0850, AUC= 0.9969, MSE=0.0226). These results are comparable to those produced using the 4666 SNPs extracted using logistic regression and p-value ($5*10^{-3}$). The worst results were (Sens=0.9562, Spec=0.8780, Gini=0.9490, Log Loss=0.5701, AUC= 0.9745, MSE=0.2010) when 50 features were used. Nonetheless, the results were significantly better than using the SNPs generated using logistic regression and a p-value of $5*10^{-5}$ (51 SNPs - Sens=0.9666, Spec=0.5411, Gini=0.6709, Log Loss=0.5581, AUC= 0.8354, MSE=0.1913). The Sensitivity value was slightly lower. However, Specificity increased by 34\%, Gini by 28\%, while LogLoss remained broadly the same. The AUC increased by 14\% and the MSE was slightly less. The results when the input set was compressed to 1000 features produced comparable results to those when logistic regression was used with a p-value threshold of $5*10^{-3}$ (4666 SNPs).   
\par
This paper places a strong emphasis on classification tasks using latent information extracted from high-dimensional genomic data, which in the present context corresponds to whether a mother will have a normal or premature delivery. This is clearly important for mitigating risk to the mother and unborn foetus. Furthermore, it provides a new and viable way to capture epistatic interactions between SNPs. However, from an extensive literature review the extraction of identified patterns from deep networks and the classification of phenotypes using GWAS data have received little attention within the research community.  
\par
One possible reason could be directly attributed to the fact that deep learning models are difficult to interpret \cite{Lipton2016}. Compressing the input set to 50 nodes shows reasonably good predictive capacity. However, it is difficult to identify which of the 4666 SNPs contribute to those 50 features. Consequently, DL approaches are characterised as black boxes where it becomes difficult to explain good results or modify models to address misclassification issues. Therefore, it would be advantageous to combine the strengths of symbolic analysis and the strengths of machine learning to create a robust method for interpreting deep learning networks.   
\par
Nonetheless, the findings in this paper do demonstrate that a GWAS classification system could provide an early screening tool for the identification of women with a genetic disposition to preterm birth. This would eventually lead to an automated, therapeutic intervention to reduce morbidity and mortality associated with preterm deliveries, and help direct medical attention toward high-risk pregnant mothers. The current protocol used by obstetricians and midwives in perinatal care does not routinely include genetic screening. Decisions are made using cardiotocography interpretation, which is well known for inter and intra-variability \cite{Fergus2017}. The protocol has a 30\% predictive capacity and is only utilized when a risk is identified. Adopting the proposed methodology in this paper would act as an early screening intervention and enhance existing perinatal care to include GWAS classification, based on high-risk loci, which would function alongside existing procedures. Such an advanced warning system would allow sufficient time for intervention.   
\section{Conclusion}
We present a novel framework for the classification of preterm birth from GWAS case-control data. We apply deep learning stacked autoencoders and classification modelling using SNP genomic data in a predominantly African-America population and generate compressed epistatic information for mothers who had term and preterm deliveries. Using our data set of 1,567 pregnant mothers, we achieve classification results (Sen=0.9562, Spec=0.8780, Gini=0.9490, Logloss=0.5901, AUC=0.9745, MSE=0.2010) using 50 hidden nodes. Minimizing the MSE below 10\% we achieved classification results (Sen=0.9289, Spec=0.9591, Gini=0.9651, Logloss=0.3080, AUC=0.9825, MSE=0.0942) using 500 nodes. Figure 3 e and f show that there is no significant evidence of overfitting when comparing the training and validation data sets. Figure 4 c also demonstrates that our framework has good predictive capacity.   
\par
These results are encouraging. However, the study needs further research to find more sophisticated strategies for mapping SNP inputs to hidden layer nodes. SNPs are symbolic, and they mean something in the context of GWAS analysis. The minute non-linear transformations of the input space occur it is very difficult to trace the amount of variance they contribute to the phenotype. This is a common problem in neural network modelling that seriously hinders genomic analysis.   
\par
In future work, we will look at several alternative extensions to this work. It may be interesting to model the SNPs of mothers who deliver term only and implement anomaly detection using autoencoders \cite{Zhou2017} to identify pregnant mothers with genetic differences - they do not necessarily have to deliver prematurely. This would provide clear groupings and act as a basis for more in-depth analysis of these genomic differences. Another interesting approach is to train an interpretable model, such as a random forest, alongside our stacked autoencoder to see, what SNPs are being used by the neural network during model construction \cite{Lipton2016}. Structured logic rules are another technique that has been proposed to reduce uninterpretability of neural networking models \cite{Hu2016}.     
\par
Overall, the proposed methodology in this paper provides a body of work that highlights the benefits of using stacked autoencoders and deep learning to detect epistatic interactions between SNPs in higher-order genomic sequences. This contributes to the computational biology and bioinformatics field, and provides new insights into the use of deep learning algorithms when analysing GWAS that warrant further investigation. To the best of our knowledge this is the first comprehensive study of its kind that presents a framework consisting of stacked autoencoders and DL classification models in GWAS analysis.   

\section*{Acknowledgement}
The dataset(s) used for the analyses described in this manuscript were obtained from the database of Genotype and Phenotype (dbGaP) found at http://www.ncbi.nlm.nih.gov/gap through dbGaP accession number phs000332.v3.p2. Samples and associated phenotype data for the CIDR Preterm Birth Boston Cohort were provided by Xiaobin Wang, M.D.
\bibliography{litrev}
\bibliographystyle{ieeetr}

\vspace{-20mm}
\begin{IEEEbiography}[{\includegraphics[width=1in,height=1.25in,clip,keepaspectratio]{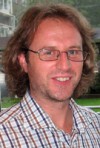}}]{Dr Paul Fergus}
 is a Reader (Associate Professor) in Machine Learning. He is the Head of the Data Science Research Centre. Dr Fergus's main research interests include machine learning for detecting and predicting preterm births. He is also interested in the detection of foetal hypoxia, electroencephalogram seizure classification and bioinformatics (polygenetic obesity, Type II diabetes and multiple sclerosis). He is also currently conducting research with Mersey Care NHS Foundation Trust looking at the use of smart meters to detect activities of daily living in people living alone with Dementia by monitoring the use of home appliances to model habitual behaviours for early intervention practices and safe independent living at home. He has competitively won external grants to support his research from HEFCE, Royal Academy of Engineering, Innovate UK, Knowledge Transfer Partnership, North West Regional Innovation Fund and Bupa. He has published over 200 peer-reviewed papers in these areas.
\end{IEEEbiography}

\vspace{-20mm}
\begin{IEEEbiography}[{\includegraphics[width=1in,height=1.25in,clip,keepaspectratio]{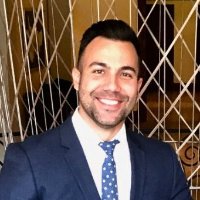}}]{Casimiro Curbelo Monta\~{n}ez}
is a PhD candidate of the Applied Computing Research Group at Liverpool John Moores University (LJMU), UK, under the supervision of Dr. Paul Fergus. He received his B.Eng. in Telecommunications in 2011 from Alfonso X el Sabio University, Madrid (Spain). In 2014, Casimiro Aday obtained an MSc in Wireless and Mobile Computing from Liverpool John Moores University. His research interests include various aspects of data science, machine learning and their applications in Bioinformatics”
\end{IEEEbiography}

\vspace{-40mm}
\begin{IEEEbiography}[{\includegraphics[width=1in,height=1.25in,clip,keepaspectratio]{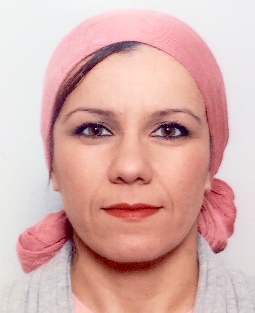}}]{Basma Abdulaimma}
received a BSc (Hons) in Computer Science from Baghdad Technology University, Iraq in 1999, an MSc in Computing and Information Systems from Liverpool John Moores University (LJMU), UK in 2013. She is currently a PhD candidate at Liverpool John Moores University. Her research interests include data science, machine learning, and artificial intelligence. She is especially interested in bioinformatics and computational biology at a molecular level particularly genetics. 
\end{IEEEbiography}

\vspace{-40mm}
\begin{IEEEbiography}[{\includegraphics[width=1in,height=1.25in,clip,keepaspectratio]{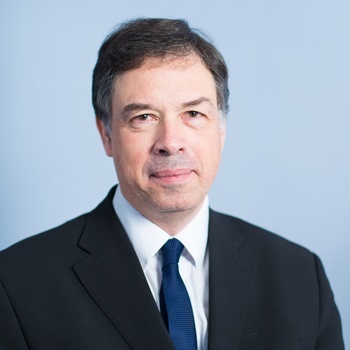}}]{Prof. Paulo Lisboa}
is Professor and Head of Department of Applied Mathematics at Liverpool John Moores University. His research focus is advanced data analysis for decision support. He has applied data science to personalised medicine, public health, sports analytics and digital marketing. In particular, he has an interest in rigorous methods for interpreting complex models with data structures that can be validated by domain experts. He is vice-chair of the Horizon2020 Advisory Group for Societal Challenge 1: Health, Demographic Change and Wellbeing, providing scientific advice to one of the world’s largest coordinated research programmes in health. A member of Council for the Institute of Mathematics and its Applications, he is past chair of the Medical Data Analysis Task Force in the Data Mining Technical Committee of the IEEE, chair of the JA Lodge Prize Committee and chair of the Healthcare Technologies Professional Network in the Institution of Engineering and Technology. He is on the Advisory Group of Performance.Lab at Prozone and has editorial and peer review roles in a number of journals and research funding bodies including EPSRC. Paulo Lisboa studied mathematical physics at Liverpool University where he took a PhD in theoretical particle physics in 1983. He was appointed to the chair of Industrial Mathematics at Liverpool John Moores University in 1996 and Head of Graduate School in 2002.
\end{IEEEbiography}

\vspace{-40mm}
\begin{IEEEbiography}[{\includegraphics[width=1in,height=1.25in,clip,keepaspectratio]{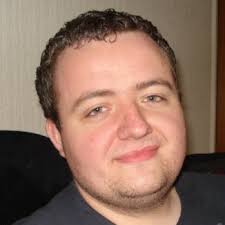}}]{Dr Carl Chalmers}
is a Senior Lecturer in the Department of Computer Science at Liverpool John Moores University. Dr Chalmers’s main research interests include the advanced metering infrastructure, smart technologies, ambient assistive living, machine learning, high performance computing, cloud computing and data visualisation. His current research area focuses on remote patient monitoring and ICT-based healthcare. He is currently leading a three-year project on smart energy data and dementia in collaboration with Mersey Care NHS Trust. As part of the project a six month patient trial is underway within the NHS with future trials planned. The current trail involves monitoring and modelling the behaviour of dementia patients to facilitate safe independent living. In addition he is also working in the area of high performance computing and cloud computing to support and improve existing machine learning approaches, while facilitating application integration. 
\end{IEEEbiography}

\end{document}